\documentclass [12pt]{revtex4}
\usepackage[english]{babel} %
\usepackage {hyperref}
\usepackage {amsmath,amssymb,caption,fancyhdr,graphicx,lineno,remreset,sidecap,url}
\usepackage{amssymb}
\usepackage{amsmath}
\usepackage{multirow}
\usepackage{minitoc}

\begin{document}
\title{Markov Chains application to the financial-economic time series prediction}

\author{Vladimir Soloviev}% V., Prof. Dr. Sc.}
\email{vnsoloviev@rambler.ru}
\affiliation{Cherkasy National University named after B. Khmelnitsky, Cherkassy, Ukraine}
% сколько букв ss в Cherkasy?

\author{Vladimir Saptsin}% V., PhD.}
\email{saptsin@sat.poltava.ua}
\affiliation{Kremenchug National University named after M. Ostrogradskii, Kremenchuk, Ukraine}

\author{Dmitry Chabanenko}% D.}
\email{chdn6026@mail.ru}
\affiliation{Cherkasy National University named after B. Khmelnitsky, Ukraine}

%\date{\today}

\begin{abstract}
In this research the technology of complex Markov chains is applied to predict financial time series. The main distinction of complex or high-order Markov Chains and simple first-order ones is the existing of aftereffect or memory. The technology proposes prediction with the hierarchy of time discretization intervals and splicing procedure for the prediction results at the different frequency levels to the single prediction output time series. The hierarchy of time discretizations gives a possibility to use fractal properties of the given time series to make prediction on the different frequencies of the series. The prediction results for world's stock market indices is presented.
\end{abstract}


%\dominitoc
%\minitoc

%\tableofcontents

\maketitle
%\section*{}
\label{contents}
\tableofcontents

\section{Introduction}
\label{sec:Intro}

Successful modeling and prediction of processes peculiar to complex systems, such as ecological, social, and economical (ESE) ones, remain one of the most relevant problems as applied to the whole complex of natural, human and social sciences (\cite{r001SamarskyMihaylov01,IvakhnenkoMGUA68, r002BogoboyashyKurbanov04,AndersenGluzmanSornette2000,PincakCurrencyStringTheory2011,PincakStringPrediction2011}).

The diversity of possible approaches to modeling such systems and, usually, more than modest success in the dynamics prediction, compel us to look for the reasons of failure, finding them not only in details, but also in the axiomatics, which relates to problem statement, chosen modeling methods, results interpretation, connections with other scientific directions.

With the appearance of quantum mechanics and relativity theory in early twentieth century new philosophical ideas on physical values, measuring procedures and system state have been established, the ones that are completely different from Newtonian notions \cite{r003ElutinKrivchenko76, r004LandauQuantNerelyativistic}.

For more than 70 years basic concepts of classical and neoclassical economic theories have been discussed by leading scientists, generating new approaches \cite{r005SapirEconTheoryNeodnorSyst}. The general systems theory has acquired recognition in the middle of the 20th century giving way to development of the new, systemic, emergent, and quantum in essence approach to investigation of complex objects, which postulates the limited nature of any kind of modeling and is based upon fixed and closed system of axioms \cite{r006BertalanfyGenSyst62}.

However, the development of this new philosophical basis of ESE systems modeling is still accompanied with numerous difficulties, and new principles are often merely declared.

Current research is devoted to investigation and application of the new modeling and prediction technology, suggested in \cite{r007KurbanovSaptsin07, r008SaptsinMarkovPaper09_}, based on concepts of determined chaos, complex Markov chains and hierarchic (in terms of time scale) organization of calculating procedures.

%\textbf{2. Analysis of prominent publications relevant to the subject}
\section{Analysis of prominent publications relevant to the subject}
\label{sec:AnPublications}

Prediction of financial-economic time series is an extremely urgent task. Modern approaches to the problem can be characterized by the following directions: 1) approximation of a time series using an analytical function and extrapolation of the derived function towards future -- so-called trend models \cite{r009LukashinAdapt03}; 2) investigation of the possible influence various factors might have on the index, which is being predicted, as well as development of econometric or more complicated models using the Group Method of Data Handling (GMDH) \cite{IvakhnenkoMGUA68,r010ZaychenkoMonogr08}; 3) modeling future prices as the decisions-making results using neuronal networks, genetic algorithms, fuzzy sets \cite{r010ZaychenkoMonogr08, r011EzhovShumsky98, r012Zaencev}. 

Unfortunately, these techniques don't produce stable forecasts, what can be explained by complexity of the investigated systems, constant changes in their structure. Although we are trying to join these directions in one algorithm, it is the latter option that we prefer, with it consisting in creating a model adequate to the process generating a price time series \cite{r013ChabM10_}. This very approach gives a chance to approach the complexity of the system, which generates the observed series, develop the model and use its properties as the prognosis.

%\textbf{3. Aims of the paper, problem statement}
\section{Aims of the paper, problem statement}
\label{sec:Aim}

Assume the time series is set by a sequence of discrete levels with constant step of time sampling  $\Delta t$. We need to generate variants of the time series continuation (prognosis scenarios) according to the relations between the sequences of absolute and relative changes discovered with the help of complex Markov chains. 

%\textbf{4. Classical modeling problems of ESE systems dynamics } % где абревиатура появилась?
\section{Classical modeling problems of ESE systems dynamics }
\label{sec:Classical}

Another peculiar feature of ESE systems, apart from complexity, is a memory, including the long-term one, as well as nonlinear and unstable nature of interactions and components, which makes it harder to predict their future behavior.

Unfortunately, mathematical models based on differential equations have no memory (there is no aftereffect), while for models with memory, where integral interrelations are used, it is not always possible to take into account nonlinearity (the integration procedure is linear by definition).

In reality, in the Cauchy problem future systems behavior is defined by its initial state and doesn't depend on the way the system reached its current state. However, it is hardly true that future behavior of a real socio-economic or socio-ecological system can be predicted by giving an immediate time ``slice'' of a variables set that describe its state. 

Let us consider possible ways to take into account past events while modeling ESE systems' dynamics, which goes beyond the boundaries of classical differential and integral equations.

Functional differential lagging equation can serve as a simple example of the dynamic model with memory, where present time is defined by the state variable $x(t)$ and depends on the past state $x(t-\tau)$ with constant \textit{time lag} $\tau=const$:
\begin{equation}
\label{eq:eq01}
x(t)=f\left(x(t-\tau)\right); t \ge t_0,
\end{equation}
where $f(x)$ is the known function, with \textit{initial conditions} being set for the half-interval $t_0-\tau \le t < t_0$ by the function $\phi(t)$:
\begin{equation}
\label{eq:eq02}
x(t)=\phi(t); t_0-\tau \le t < t_0.
\end{equation}
Given the \ref{eq:eq02} equation \ref{eq:eq01} has the only solution, defined by \textit{recurrent ratios}:
\begin{equation}
   \label{eq:eq03}
x(t)=
\begin{cases}
f\left(\phi(t-\tau)\right);\text{ if } t_0 \le t <t+\tau;\\
f\left(f\left(\phi(t-\tau)\right)\right);\text{ if } t_0 +\tau \le t <t+2\tau;\\
f\left(\left(f\left(\phi(t-\tau)\right)\right)\right);\text{ if } t_0 +2\tau \le t <t+3\tau;\\
\cdots
\end{cases}
\end{equation}

Using \textit{Dirac delta function}, as defined by ratios:
\begin{equation}
\label{eq:eq04}
\delta(t)=0, \text{ if } x \ne 0; \int\limits_{-\infty}^{+\infty}{\delta(t)dt}=1,
\end{equation}
we can formally rewrite equation \ref{eq:eq01} in the integral form:
\begin{equation}
\label{eq:eq05}
x(t)=\int\limits_{-\infty}^{t}{dt_1 f\left(x(t_1)\right)H(t_1,t)}; H(t_1,t) \equiv \delta \left(t_1 - (t-\tau)\right); t \ge t_0.
\end{equation}
Delta function is not a function in the conventional interpretation and is related to the class of generalized functions that were mathematically described only in the middle of the last century \cite{r014SobloevFunctProstr} (physics started using this function much earlier). Its classical form is considered to be a limit of the ``peak'' sequence, with its centre set in the point of origin. The afore-mentioned ``peaks'' indefinitely converge widthway, indefinitely increase throughout the height and have a unit area.

An approximate classic integral analogue of the equation \ref{eq:eq05} can be derived by substituting $\delta(t)$ with an ordinary function - some specific narrow enough ``peak'' of a unit area, a certain finite width $\backsim \Delta t$ as well as a finite height . The derivative of the \textit{Fermi function}  is one of the possible examples:
\begin{equation}
\label{eq:eq06}
\Phi(t)=\frac{1}{1+exp\left(\frac{-t}{\theta}\right)};\delta(t) \approx \frac{1}{\theta\left(2+exp\left(\frac{-t}{\theta}\right)+exp\left(\frac{t}{\theta}\right)\right)}.
\end{equation}
If the system's state in the moment $t$, $x(t)$, is defined not by one, as in \ref{eq:eq01}, but $k$ ($k=2,3,4,\cdots$) of her past states $x(t-\tau_1), x(t-\tau_2),\cdots x(t-\tau_k)$ in the following moments of time $(t-\tau_1), (t-\tau_2), \cdots, (t-\tau_k)$ respectively $(\tau_1=const, \tau_2=const, \cdots, \tau_k=const, \tau_1 > \tau_2 > \cdots > \tau_k > 0 )$, then instead of (\ref{eq:eq01}), (\ref{eq:eq02}), (\ref{eq:eq05}) we get:
\begin{equation}
\label{eq:eq07}
x(t)=f\left(x(t-\tau_1);x(t-\tau_2);\cdots;x(t-\tau_k)\right);t \ge t_0;
\end{equation}
\begin{equation}
\label{eq:eq08}
x(t)=\phi(t); t_0-\tau_1 \le t < t_0;
\end{equation}
\begin{equation}
\label{eq:eq09} 
\begin{array}{c}
x(t)=\int\limits_{-\infty}^{t}{dt_1}\int\limits_{-\infty}^{t}{dt_2}\cdots\int\limits_{-\infty}^{t}{dt_kf\left(x(t_1),x(t_2),\cdots,x(t_k)\right)}.\\
\delta\left((t_1-(t-\tau_1)\right)\delta\left((t_2-(t-\tau_2)\right)\cdots\delta\left((t_k-(t-\tau_k)\right); t \ge t_0.
\end{array}
\end{equation}

Therefore if the system's state in the time moment $t$ depends on the infinite sequence of its past states, the integral analogue of the functional differential lagging equation will, generally speaking, contain an integral of the infinite multiplicity. At the same time the infinite amount of past states can relate to both finite $(t-\tau_1;t)$ (short-term memory) and infinite $(-\infty; t)$ (long-term memory) time span.

Pay attention that the classic integral lagging equation is one of the Volterra type \cite{r015HeliFunctDiffury}:
\begin{equation}
\label{eq:eq10}
x(t)=\int\limits_{-\infty}^{t}{F\left(x(\tilde{t});t;\tilde{t}\right)d\tilde{t}},
\end{equation}

where $F\left(x(\tilde{t});t;\tilde{t}\right)$ - is an arbitrary (generally nonlinear) function of variables $x(\tilde{t});t;\tilde{t}$, which allows to take into account system's memory of its past states only in the additive approximation, which becomes evident, if the right section \ref{eq:eq10} is rewritten in the following way:

\begin{equation}
\begin{array}{c}
\int\limits_{-\infty}^{t}{F\left(x(\tilde{t});t;\tilde{t}\right)d\tilde{t}} \equiv 
\int\limits_{t_1}^{t}{F\left(x(\tilde{t});t;\tilde{t}\right)d\tilde{t}}+
\int\limits_{t_2}^{t_1}{F\left(x(\tilde{t});t;\tilde{t}\right)d\tilde{t}}+\cdots = \\
F\left(x(\tilde{t_1});t;\tilde{t_1}\right) \cdot (t-t_1)+ F\left(x(\tilde{t_2});t;\tilde{t_2}\right)\cdot(t_1-t_2)+ \cdots;\\
t>t_1>t_2>\cdots; \tilde{t_1} \in \left[t_1,t\right]; \tilde{t_2} \in \left[t_2,t_1\right];\cdots
\end{array}
\end{equation}

In connection with it note that the equation \ref{eq:eq09} in case of an additive dependency of contemporaneity on the past, i.e. in case:
\begin{equation}
\label{eq:eq12}
f\left(x(t-\tau_1);x(t-\tau_2);\cdots\right) \equiv f_1\left(x(t-\tau_1)\right);f_2\left(x(t-\tau_2))\right);\cdots
\end{equation}
becomes a particular case of the equation \ref{eq:eq10} with the following integrand:
\begin{equation}
\label{eq:eq13}
F\left(x(\tilde{t});t;\tilde{t}\right) \equiv f_1\left(x(\tilde{t})\right)\delta\left(\tilde{t}-(t-\tau_1)\right)+f_2\left(x(\tilde{t})\right)\delta\left(\tilde{t}-(t-\tau_2)\right)+\cdots
\end{equation}

Meaningful analysis of nonlinear models dynamics with memory, in which the future is defined by the infinite amount of states in the past is generally possible only in case of a discrete representation. The results of such analysis will be approximated, i.e. will contain uncertainty, which has to be considered \textit{endogenous}, i.e. internal, and peculiar to this very system.

With a certain level of time sampling, models with memory both \ref{eq:eq07} and \ref{eq:eq10} becomes:

\begin{equation}
\label{eq:eq14}
x(n+1)=f\left(x(n);x(n-1);x(n-2)\cdots\right).
\end{equation}

To take into account and quantify the uncertainties, observed in ESE as well as other complex systems probability models are normally used. However their application is based on doubtful hypotheses, while the statistical interpretation of the results is not always informative enough and results might not correspond with the real process occurring within the system. In particular, the well-known problem of $1/f$--noise (look for example \cite{r016BukingemShumy}), closely connected to the presence of long-term memory in complex systems, implies the absence of the mean temporary value (as a limit of a certain time span converging to infinity, which serves as the basis for averaging) for any process occurring in such kind of system. Therefore such processes can't have a rigorous statistical substantiation.

%\textbf{5. Modern concepts in ESE systems modeling}
\section{Modern concepts in ESE systems modeling}
\label{sec:Modern}

New approaches to modeling and prediction of complex nonlinear systems dynamics with memory are based on the use of determined chaos and neural networks technologies (cf. e.g. \cite{r011EzhovShumsky98, r017LorenzNonlinear89, r018PetersChaosOrder}). Both investigation and realization of such techniques has become possible only with the appearance of quick-operating computers. Use of the recurrent computational process has become the general feature for all these technologies:
\begin{equation}
\label{eq:eq15}
x_{n+1}=f_n\left(f_{n-1}\left(\cdots\left(f_1(x_1)\cdots\right)\right)\right), n=1,2,\cdots,
\end{equation}

where $f_{i} (x_{i} )$ is a certain nonlinear mapping of a multi-dimensional vector $x_{i} $, $i$ - discrete, real or fictitious, time. Identification of the model \ref{eq:eq15} is reduced to the determination of functions $f_{i} (x_{i} )$, while the differences between the models of determined chaos and neural networks are connected with the function type and methods of its definition (neural network models normally use rather narrow class of $f_{i} (x_{i} )$ mappings \cite{r012Zaencev}). Generally speaking, stability or convergence of the process \ref{eq:eq15} is not required, whereas a single-step set of vector $x_{i} $ components as well as their time dynamics can be of great interest.

For the particular case of the model \ref{eq:eq15}, introducing corresponding lagged variables, a model \ref{eq:eq14} can be transformed.

Both determined stable processes, described by integro-differential equations, and random processes, which also include complex Markov chains (CMC), can be formally considered as separate extreme cases of determined chaos models realization \ref{eq:eq15}. Given the sampling scale, which tends to zero, if such a tendency makes sense and corresponding limits exist, we derive classical differential and integral problem statement. Finite $\Delta t$ allows to get models with discrete time, which in the general case in the corresponding phase space (which also includes lagged variables) can produce both measurable sets (discrete or continuous) that allow probabilistic interpretation and  those of the special structure -- fractals \cite{r019FederFractals}, that can't be always interpreted in that way.

Various digital generators of so-called random sequences used in imitational modeling can be an example of determined chaos models that allow probabilistic interpretation.

Let us note that in reality there are no accurate procedures that would give an opportunity to distinguish a ``real'' random sequence from the pseudorandom one.

%\textbf{6.  Markov chains prediction technology}
\section{Markov chains prediction technology}
\label{sec:Markov}

Suppose there is a sequence of a certain system discrete states. From this sequence we can determine transitions probabilities between the two states. Simple Markov chain is a random process, in which the next state probability depends solely on the previous state and is independent from the rest of them. Complex Markov chain, unlike the simple one, stands for the random process, in which the next state probability depends not only on the current, but also on the sequence of several previous states (history). The amount of states in history is the order of the Markov chain.

Theory of simple Markov chains is widely presented in literature, for example \cite{r020TihonovMironovMarkovProcesses}. As for the high order Markov chains, modern literature \cite{r021KornVMSpravochnik73} can offer us a mere definition. 
Developing complex or high order Markov chain's properties is not widely presented in modern scientific publications. It's necessary to mention the papers \cite{RafteryHighOrderMarkovChain, RafteryTavare94} where properties of complex Markov Chains are developed, but no prediction algorithm is proposed there. The development of prediction method, based on complex Markov chains, is proposed in this paper.

Markov chain of the higher order can be brought to a simple Markov chain by introducing the notion of a ``generalized state'' and including a series of consequent system's states into it. In this case, tools of simple Markov chains can be applied to the complex ones.

Investigated dynamic series is a result of a certain process. It is assumed that this process is determined, which implies the existence of a causal dependence of further states on history. It is impossible to fix and analyze the infinite history, which puts obstacles in the way of an accurate detection of this influence and making precise predictions.

The problem consists in the maximal use of information, which is contained in the known segment of the time series, and subsequent modeling of the most probable future dynamics scenario. 

The observed process is described as a time series of prices $p_t$ with the given sampling time span $\Delta t$
\begin{equation}
\label{eq:eq16}
 p_{ti}=p(t_0 + i\Delta t). 
\end{equation}
Discrete presentation of the time series is in fact a way of existence of this very system. New prices are formed on the basis of contracts or deals, made on the market in certain discrete moments of time, while the price time series is a series of the averaged price levels during the chosen time intervals. While making a decision each trader, who is an active part of the pricing system, works solely with discrete series of the chosen time interval (e.g. minute, 5-minute, hourly, daily etc.). For $\Delta t \to 0$ the accuracy of data presentations reaches a certain limit, since for relatively small $\Delta t$ the price leaps in the moment of deal, while staying unchanged and equal to the last deal during the time between the two deals. Hence, the discreteness of time series has to be understood not only as a limited presentation of activity of the complex financial system, but also as one of the principles of its operation \cite{r007KurbanovSaptsin07, r008SaptsinMarkovPaper09_, r022SaptsinSoloviev_,r022SapSolArxiv_,r023SolDerbMonogr2010_}.

The time series of initial conditions has to be turned into a sequence of discrete states. Let us denote the amount of chosen states as $s$, each of them being connected to the change in the quantity of the initial signal (returns). For example, consider the classification with two states, first of which corresponds to positive returns as the price increases, while the second one -- to negative as it descends. Generally all possible increments of the initial time series are divided into $s$ groups. Ways of division will be discussed further. 

Next we develop predictions for the time series of sampled states. For the given order of the Markov chain and the last generalized state the most probable state is chosen to be the next one. In case if ambiguity occurs while the state of maximum probability is being evaluated, an algorithm is used that allows reducing the amount of possible prediction scenarios. Therefore we get the series of predicted states that can be turned into a sampled sequence of prognostic values.

Evaluation of increments, prediction, and subsequent restoration are conducted for the given hierarchy of time increments $t$. To use the given information as effectively as possible, the prediction is conducted for time increments $t=1,2,4,8,...$, or a more complex hierarchy of increments and subsequent ``splicing'' of the results derived from different prediction samplings.

The procedure of prediction and splicing is iterative and conducted starting from smaller increments, adding a prediction with the bigger time increment on every step.

As the sampling time step $t$ increases, the statistics for the investigation of Markov chains decreases, whereas the biggest sampling step, which takes part in the prognostication, limits itself. To supplement the prediction with the low-frequency component the approximation of zero order is being used in the form of a linear trend or a combination of a linear trend and harmonic oscillations \cite{r024SapChFourierKharkov, r025ChabS10_}. 

%\textbf{7. Prediction construction algorithm}
\section{Prediction construction algorithm}
\label{sec:PredConstAlg}

Let us consider the consequence of operations, required for the prognostic time series construction. To do this we need to set the following parameters:

1) The type of time increments hierarchy (simple -- powers of two, complex -- product of powers of the first simple numbers).

2) Values of $s$ -- the amount of states and $r$ -- the order of the Markov chain. These parameters can be individual for every sampling level; finding of optimal parameters is done experimentally. 

3) Threshold values $\delta$, and minimal number of transitions $N_{min}$.

Prediction construction algorithm includes the following steps: 

1) Generating hierarchy of time increments - $t$ sequence. The maximal of them has to correspond to the length of a prognostic interval $N_{max}$.

2) For every time increment $\Delta t$, as the increments increase, a prediction of states and restoration of the time series along the prognostic states is conducted. Current stage includes following actions: 

2.1. Evaluating increments (returns) of the series with $\Delta t$ sampling.

2.2. Transforming the time series of increments into the series of state numbers ($1..s$).

2.3. Calculating transition probabilities for generalized states. 

2.4. Constructing the series of prognostic states using the procedure of defining the most probable next state. 

2.5. Restoring the value series from the state series with $\Delta t$ sampling. 

2.6. Splicing the prediction of $\Delta t$ sampling with the time series derived from splicing of the previous layers (with the lesser step $\Delta t$). In case if the current time series is the first one, the unchanged time series will come as a result of splicing. 

3) To splice the last spliced time series with the continuation of the linear trend, created along all previously known points.

The time series, spliced with the linear trend, is the result of prediction. Let us consider the stages of the given algorithm in detail.

%\subsection{8. States in complex Markov chains and approaches for defining them }
\section{States in complex Markov chains and approaches for defining them }
\label{sec:States}

In everything that concerns current technology, states are connected to the measuring of a prognostic value. There is a number of ways to classify returns in states, from which the following are suggested. One of them is the classification based on the homogeneity principle as concerning the amount of representatives in classes; based on the homogeneity principle of deviation, as well their combinations for different deviation modules. 

Increment or returns of the time series serves as the basis for states classification \cite{r025ChabS10_, r026solovievmatheconomics_}. Absolute $r_{a} $ and relative $r_{t}$ increments of the time series are considered:
\begin{equation}
 \label{eq:eq17}
r_a = p_t - p_{t-\Delta t} , 
\end{equation}
\begin{equation}
 \label{eq:eq18)}
 r_{t} =\frac{p_{t} -p_{t-\Delta t} }{p_{t}} ,
\end{equation} 

where $p_t$ -- is the input time series of price dynamics, $\Delta t$ -- sampling interval, which is chosen for subsequent analysis. It is known that mathematical expectation of the returns time series equals zero, whereas variation comes as the measure of time series volatility. Based on returns values $r_t$ classification and transformation of values to the time series of discrete states are conducted. One of the classification principles  is homogeneity according to the amount of class representatives. This classification divides the set of all increments into $s$ groups equal in number. Calculated with the given sampling, time series increments are then systematized in growth and divided into equal parts. Thus we define limit values $\{r_{lim,i}\}$, which are used afterwards during transformation of the returns into class numbers. Large number of identical states can cause certain problems, such as identical bounds of several neighbouring states. It creates a number of states with no representatives, which makes correction of the division a necessary action. In that way we will reach the largest possible homogeneity in state division. Classification is conducted along the following algorithm \cite{r027SolSapChDrezden09, r028ChabRiga10}:

\begin{equation}
\label{eq:eq19}
s_t=(i \mid r_{lim,i-1} > r_t > r_{lim,i})
\end{equation}
where $s_t$ is the number of state, which corresponds to the moment of time $t$, for which the returns level was computed $r_t$; i is the number of state $[1\dots s]$, which is characterized by the interval $[r_{lim,i-1},r_{lim,i}]$ corresponding to the calculated returns level $r_t$. 

Apart from the returns interval, given by the aforementioned values $[r_{lim,i-1},r_{lim,i}]$, a mean returns value is chosen for every state $r_{avg,i}$, which will be used in time series values transformation according to the prognostic discrete states.

Another way of dividing the time series into states implies dividing the interval of returns values into equal parts, from minimal to maximal deviation. In this case homogeneity according to the amount of representatives in states does not occur. In fact this method differs from the previous one in terms of defining limit values $\{r_{lim,i}\}$. Possible combined ways of division, in case of which the limit value, dependent on standard deviation, is used instead of maximal and minimal value, and division is conducted homogenously according to the deviation. 

Since the real causal dependence is unknown during the process, to find it adequate state classification, which would allow to reveal vital dependencies of the time series, is required. We suggest a couple of ways to divide the time series into states, which in the first place allow to preserve adequate transition probabilities between states, as well as prevent averaged deviations inside the states from affecting the accuracy of the derived prediction. 

To check the efficiency of division we conduct the sampling procedure and classify the increments according to each hierarchy. Having completed that, we restore the time series using known states for each hierarchy and finish the splicing procedure. Since the state series correspond to the initial time series, we get the curve, with deviation, caused exclusively by the state averaging mistake (quantum mistake). Thus, having set a certain value of state numbers $s$ and carried out sampling, restoring, and sampling procedures (excluding prediction), we get absolute sampling (quantum) mistake. % не квантовая ошибка, а квантованая!!!

Increasing the number of states, we improve the accuracy of restoration, however one should remember, that the choice of the quantum levels is limited by the fact that the transition probabilities definition with sufficient accuracy is required, which is confirmed by artificial test time series prediction experiments.

%\subsection{9. Step-by-step prediction procedure. Defining the most probable state on the next step, prediction scenarios}
\section{Step-by-step prediction procedure. Defining the most probable state on the next step, prediction scenarios}
\label{sec:StepPred}

Predicting procedure uses the most probable state as the next one under current circumstances. Probability matrix of state transitions is used for the afore-mentioned purpose. In this case, you have to take into account that probabilities are calculated with a certain mistake. We cannot precisely compute the probabilities, since it is impossible to derive an infinite time series, and only a part of the time series is known -- the known part serves as the basis for probabilities. The second important aspect implies the case of several states with maximal probability. 

To prevent the omission of the states, for which the probabilities are computed with a mistake, one should add a state with maximal probability to the states, which are located in the distance of $\delta$ from the maximal one. The value of parameter $\delta$  depends on the probability evaluation mistake and requires experimental refinement.

If $\delta > 0$, the number of states with maximal probability increases in comparison to the value $\delta = 0$. Let us call a couple of neighbouring states with maximal probability a cluster. Cluster states with average deviation values are supposed to have the largest probability. To predict the dynamics, let us confine ourselves to one or two most probable states. To define them a following algorithm is suggested:

1) If levels (discretized increments) create several clusters (cluster is a group of several neighbouring levels -- cluster elements, minimal cluster is a single isolated level) with maximal probability, we choose the largest cluster. 

2) If the number of cluster elements is odd, as $k_{max}$  we choose a central element.

3) If the number of cluster elements is even, we consider two central cluster elements and choose as $k_{max}$ the one, which is closer to the centre of distribution. 

4) If two central cluster elements are equidistant to the centre of distribution, we consider both cases as possible variants of $k_{max}$ values (bifurcation point).

5) If there are several clusters of maximal size, we consider them as new elements, which can also form clusters that will undergo the same steps 1)-4).

This principle is based on the following ideas: 

1) If there are two neighbouring states of maximal probability, it is better to take the one, which is closer to the centre of distribution, in order to minimize the risk of occurrence of false linear trends in the prediction.

 2) If levels of maximal probability are not the neighbouring ones, at least two variants have to be considered, as it can be connected to the bifurcations that should not be omitted. 

3) If the prediction is carried out according to 1) (on all stages of the hierarchy), we receive a certain approximation of the lower limit of the prediction, whereas in case of 2) -- we get an approximation of the upper limit.

Hence this algorithm can adequately restore the case of possible bimodal probability distribution, it is proposed to consider 2 prediction scenarios. 

In case of the complex Markov chains, probability of the next state depends not only on the previous state, but also on the sequence of $r$ states, which have occurred before given. In this case, it is necessary to calculate transition probabilities from the sequence of $r$ states into the $r+1$ state. Formally, these probabilities can be written into the rectangular table of $(r^s, s)$ size. 

Having generalized the notion of ``present state'' and included a sequence of $r$ preceding states into it, we can reduce Markov chains of $r$ order to the chain of the first order. Thus transition probabilities can be written into rectangular matrices of $(r^s,r^s)$, that come as transition probability matrices for generalized states. 

The process of prediction implies the following: the last state is chosen (in case of Markov chains of an order $r > 1$ a sequence of $r$ latest states is taken). The probability of transition from current state to all possible states is defined. From all possible states a state with maximal probability is chosen. It is possible that several states with maximal probability occur, which can be explained by the bimodal probability distribution. The process of decision-making in this case is described later.

The chosen most probable state is taken as the next prognostic state and the procedure is repeated for the next (last added) state. Thus we receive a time series of prognostic states for the given sampling time $\Delta t$.

Further according to the received state sequence and known initial value the time series is being restored for the given time sampling $\Delta t$. In this case every state implies $\Delta t$ points of the time series. On the stage of state classification every state was connected to the average increment $r_{avg,i}$, which is added to the value of the last point in the time series, and the next discrete point is computed. Intermediate points are filled as linear interpolation of two known neighbouring points. Algorithm of $y_t$ time series values restoration according to the initial price $p_t$ and a series of average increments $r_{avg,ik}$, corresponding to the prognostic states $s_k$, can be given by a sequence of calculations:

%\begin{multline}
\begin{equation}
\label{eq:eq20}
\begin{array}{c}
y_t=p_t,\\
y_{t+1}=y_{t}+r_{avg,i1}/\Delta t= p_t+r_{avg,i1}/ \Delta t,\\
y_{t+2}=y_{t+1}+r_{avg,i1}/\Delta t=p_t+2r_{avg,i1}/\Delta t,\\
\dots \\
y_{t+\Delta t-1}=y_{t+\Delta t-2}+r_{avg,i1}/\Delta t=p_t+(\Delta t-1)r_{avg,i}/t,\\
y_{t+\Delta t}=y_{t+\Delta t-1}+r_{avg,i1}/\Delta t=p_t+\Delta t r_{avg,i}/\Delta t= p_t+ r_{avg,i1},\\
y_{t+\Delta t+1}=y_{t+\Delta t}+r_{avg,i2}/\Delta t=p_t+ r_{avg,i1}+ r_{avg,i2}/\Delta t,\\
\dots \\
y_{t+n\Delta t-1} =y_{t+n\Delta t-2} +r_{avg,in} /\Delta t=p_{t} +\sum _{k=1}^{n-1}r_{avg,ik}  +\frac{\left(\Delta t-1\right)}{\Delta t} r_{avg,ik},
\\
y_{t+n\Delta t} =y_{t+n\Delta t-1} +r_{avg,in} /\Delta t=p_{t} +\sum _{k=1}^{n}r_{avg,ik}. \\
\end{array}
\end{equation}
%\end{multline}

%y_{t+n\Delta t-1} =y_{t+n\Delta t-2} +r_{avg,in} /\Delta t=p_{t} +\sum _{k=1}^{n-1}r_{avg,ik}  +\frac{\left(\Delta t-1\right)}{\Delta t} r_{avg,ik} 

%\begin{equation} \label{eq:eq20} y_{t+n\Delta t} =y_{t+n\Delta t-1} +r_{avg,in} /\Delta t=p_{t} +\sum _{k=1}^{n}r_{avg,ik}   \end{equation} 
% multline скорее всего сам вставляет номер формулы. Поэтому переношу туда последнее уравнение и лейбл. 

%\subsection{10. Time increments hierarchy and splicing procedure}
\section{Time increments hierarchy and splicing procedure}
\label{sec:Hierarchy}

Time series increments will be computed with different steps. For example, analogous to the discrete Fourier transform, time increments are equal the powers of 2 are considered. First, we calculate increments as a remainder of two nearest neighbouring time series values, then next nearest values are considered with the step of 2, 4, 8, 16 etc. Let us mark this difference in time as $\Delta t$.

For every $\Delta t$ we conduct an increment time series transformation leading to a time series of states. Further we predict the future sequence of states and restore the time series with the given sampling rate according to the prognostic series of states.

Time series, received as a result of restoring for different $\Delta t$, undergo the splicing procedure, which gives out an actual prognostic time series. 

Thus an increment hierarchy is chosen, where each one is responsible for its own sampling rate, which serves as a basis for predicting, restoring and splicing. 

The splicing process implies the following. The procedure is iterative. With every next (along with the increasing step) sampling time the series corrects itself, driving the prediction, formed under lower $\Delta t$, to its actual point. Transformations that are conducted during splicing can be written down in the form of the following calculations. 

Suppose the splicing procedure has been finished for all time increments $\Delta t<Delta t_i$, the prediction has been done under the $\Delta t_i$ sampling according to formulae \ref{eq:eq20}, and as a result a time series $y_i$ has been derived. Let us consider the iterative splicing procedure of the received series $y_i$ with the series, acquired during all preceding splicing procedures $g_i$.

Since the series $y_i$ contains system points only in moments aliquot to $\Delta t_i$, and other points of the series are interpolated, the process of splicing implies the substitution of these interpolated points with the values of system points from previous $\Delta t < \Delta t_i$, which are contained in the series of results of previous splicing procedures $g_i$. Splicing algorithm can be written in the sequence of computations:

%\begin{multline}
\begin{equation}
\label{eq:eq21}
\begin{array}{c}
z_t=g_t=p_t,\\
z_{t+1}=g_{t+1}+\left(y_{t+\Delta t_i} - g_{t+\Delta ti}\right)/\Delta t_i,\\
z_{t+2}=g_{t+2}+2\left(y_{t+\Delta t_i} - g_{t+\Delta ti}\right)/\Delta t_i,\\
\dots \\
z_{t+\Delta t-1}=g_{t+\Delta t -1}+\left(\Delta t_i - 1\right)\left(y_{t+\Delta t_i} - g_{t+\Delta ti}\right)/\Delta t_i,\\
z_{t+\Delta t}=g_{t+\Delta t}+\left(\Delta t_i\right)\left(y_{t+\Delta t_i} - g_{t+\Delta ti}\right)/\Delta t_i = y_{t+\Delta_{ti}},\\
z_{t+\Delta t+1}=g_{t+\Delta t +1}+\left(\left(y_{t+2\Delta ti} - g_{t+2\Delta ti}\right)- \left(y_{t+\Delta ti} - g_{t+\Delta ti}\right)\right)/\Delta t_i,\\
z_{t+\Delta t+2}=g_{t+\Delta t +2}+2\left(\left(y_{t+2\Delta ti} - g_{t+2\Delta ti}\right)- \left(y_{t+\Delta ti} - g_{t+\Delta ti}\right)\right)/\Delta t_i,\\
\dots \\
z_{t+n\Delta t-1} =g_{t+n\Delta t-1} +\frac{\left(\Delta t-1\right)}{\Delta t} \left(\left(y_{t-n\Delta t} -g_{t-n\Delta t} \right)-\left(y_{t-(n-1)\Delta t} -g_{t-(n-1)\Delta t} \right)\right),\\
z_{t+n\Delta t} =g_{t+n\Delta t} +\left(\left(y_{t-n\Delta t} -g_{t+n\Delta t} \right)-\left(y_{t+(n-1)\Delta t} -g_{t+(n-1)\Delta t} \right)\right)= \\ 
=g_{t+(n-1)\Delta t} -y_{t+(n-1)\Delta t} -y_{t-n\Delta t}. 
\end{array}
\end{equation}
%\end{multline}

%\begin{equation} \label{eq:eq21} \begin{array}{l} {z_{t+n\Delta t} =g_{t+n\Delta t} +\left(\left(y_{t-n\Delta t} -g_{t+n\Delta t} \right)-\left(y_{t+(n-1)\Delta t} -g_{t+(n-1)\Delta t} \right)\right)=} \\ {=g_{t+(n-1)\Delta t} -y_{t+(n-1)\Delta t} -y_{t-n\Delta t} } \end{array} \end{equation} 

\clearpage

\section{Results of stock indices prediction}

In this section we offer the results of stock indices prediction. The stock's indices databases are available from \cite{Finance_yahoo}. Point 2000 indicates the starting moment of the prognosis: March 24, 2011. The green line on the next figures indicates real indice's or price's values. Our software for time series forecasting by the proposed methods is available from our website: \url{http://kafek.at.ua/MarkovChains1_2_20100505.rar}.

\begin{figure}[tbhp]
\begin{minipage}[h]{0.47\linewidth}
\includegraphics[width=1\linewidth]{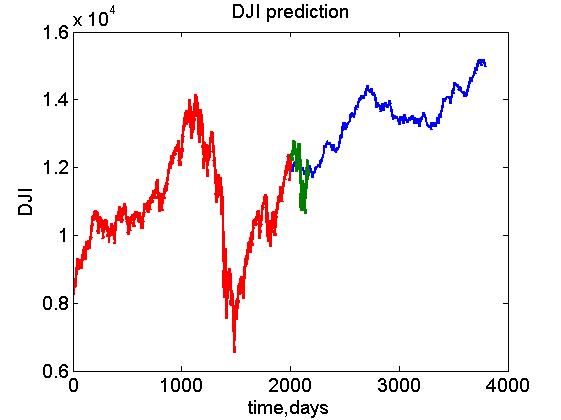} %[width=6.16in,height=4.35in]
\end{minipage}
\hfill
\begin{minipage}[h]{0.47\linewidth}
\includegraphics[width=1\linewidth]{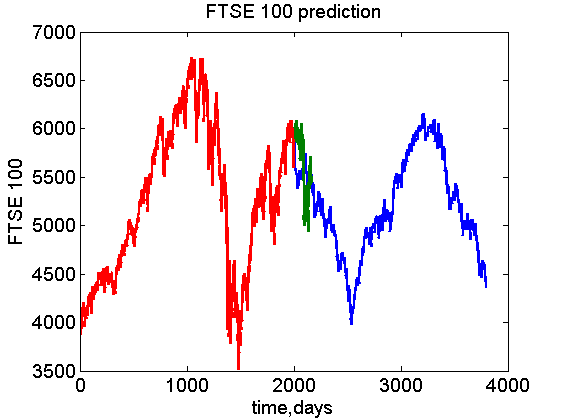} %[width=6.16in,height=4.35in]
\end{minipage}

\caption{Stock indices prediction. a) Dow Jones Industrial Average - DJI (USA). b) FTSE 100 (Great Britain)}
\label{fig:Fig01}
\end{figure}

%\begin{figure}[tbhp]
%\includegraphics[width=6.16in,height=4.35in]{02_FTSE_prognoz_itog.png}
%\caption{FTSE (Great Britain) prediction.}
%\label{fig:Fig02}
%\end{figure}
\begin{figure}[tbhp]
\begin{minipage}[h]{0.47\linewidth}
\includegraphics[width=1\linewidth]{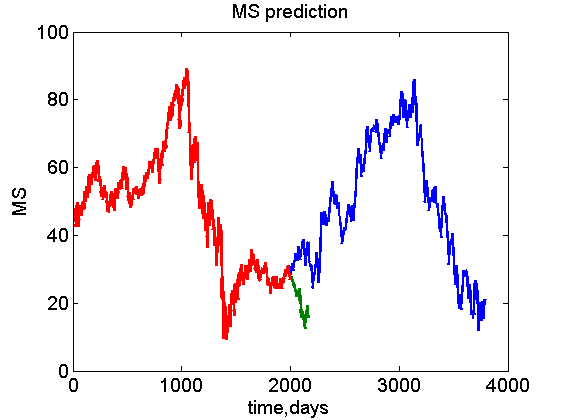} %[width=6.16in,height=4.35in]
\end{minipage}
\hfill
\begin{minipage}[h]{0.47\linewidth}
\includegraphics[width=1\linewidth]{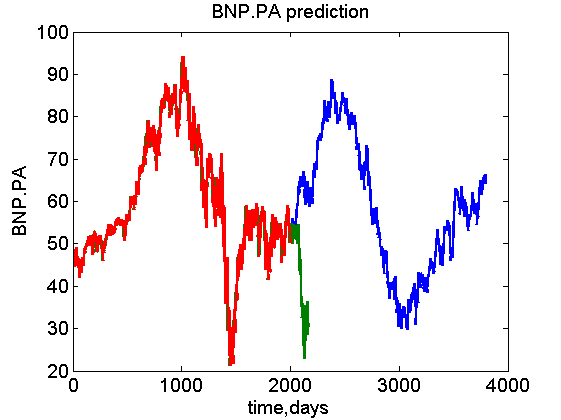} %[width=6.16in,height=4.35in]
\end{minipage}

\caption{Financial companie's share prices forecasting. a) Morgan Stanley (USA). b) BNP Paribas (France)}
\label{fig:Fig03}
\end{figure}

%\begin{figure}[tbhp]
%\includegraphics[width=6.16in,height=4.35in]{03_MS_prognoz_itog.png}
%\caption{Morgan Stanley stock prices prediction (USA).}
%\label{fig:Fig03}
%\end{figure}

%\begin{figure}[tbhp]
%\includegraphics[width=6.16in,height=4.35in]{04_BNP_PA_prognoz_itog.png}
%\caption{BNP Paribas (France) stock prices prediction.}
%\label{fig:Fig04}
%\end{figure}
\clearpage

\begin{figure}[tbhp]
\begin{minipage}[h]{0.47\linewidth}
\includegraphics[width=1\linewidth]{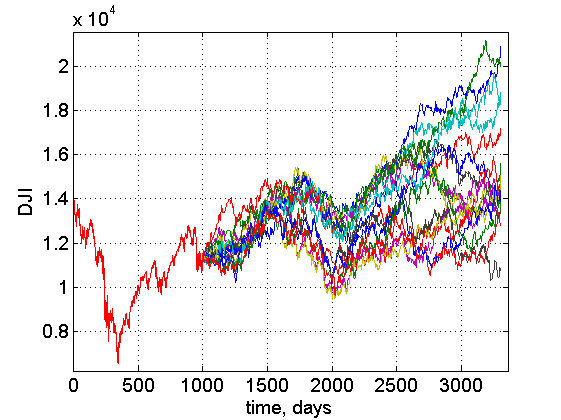} %[width=6.16in,height=4.35in]
\end{minipage}
\hfill
\begin{minipage}[h]{0.47\linewidth}
\includegraphics[width=1\linewidth]{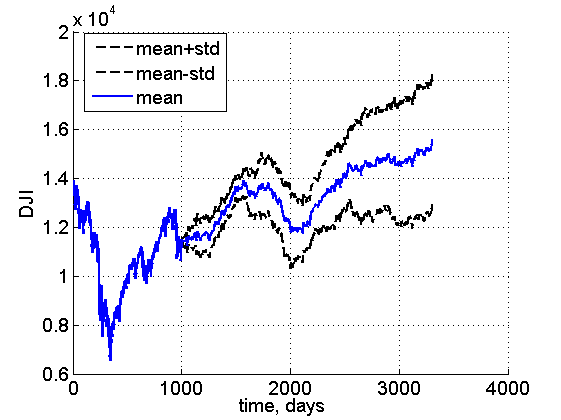} %[width=6.16in,height=4.35in]
\end{minipage}
\caption{Dow Jones Industrial Average - DJI (USA). a) Prediction series, calculated with different learning set's length. b) Mean value and standard deviation for prediction series.}
\label{fig:Fig12}
\end{figure}

\begin{figure}[tbhp]
\begin{minipage}[h]{0.47\linewidth}
\includegraphics[width=1\linewidth]{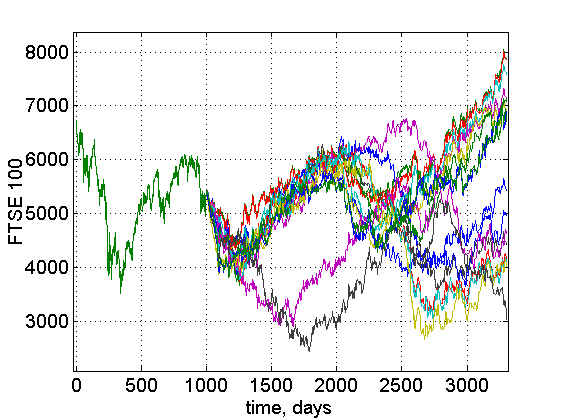} %[width=6.16in,height=4.35in]
\end{minipage}
\hfill
\begin{minipage}[h]{0.47\linewidth}
\includegraphics[width=1\linewidth]{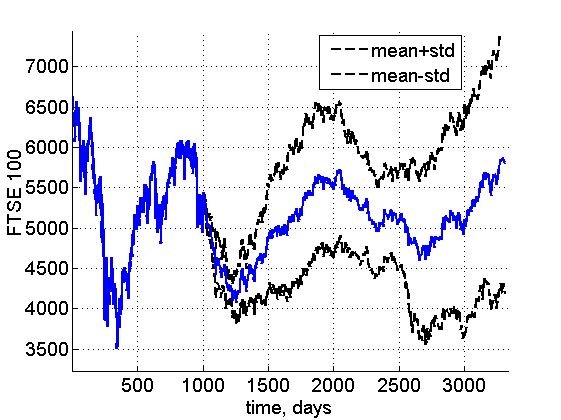} %[width=6.16in,height=4.35in]
\end{minipage}
\caption{FTSE 100 index prediction. a) Prediction series, calculated with different learning set's length. b)  Mean value and standard deviation for prediction series.}
\label{fig:Fig13}
\end{figure}

Prediction time series with different input learning set's length are shown at the fig.\ref{fig:Fig12} and \ref{fig:Fig13}. Prediction series for DJI at the figure \ref{fig:Fig12} are more correlated, than FTSE index at the figure \ref{fig:Fig13}. At the subplot b) of the above mentioned plots the mean value and standard deviations of the prediction's series are presented. The time of prediction series beginning on the next figures is the point 1000 and correspond to October 14, 2011.

%\begin{figure}
%\centerline{\includegraphics[width=6.16in,height=4.35in]{12a_dji_allprogn.png}}
%\caption{ Dow Jones Industrial average. Prediction series, calculated with different learning set's length.}
%\label{fig:Fig12a}
%\end{figure}

%\begin{figure}
%\centerline{\includegraphics[width=6.16in,height=4.35in]{12b_dji_mean.png}}
%\caption{ Dow Jones Industrial average. Mean value and standard deviation for prediction series.}
%\label{fig:Fig12b}
%\end{figure}

%\begin{figure}
%\centerline{\includegraphics[width=6.16in,height=4.35in]{13a_ftse_allprogn.png}}
%\caption{ FTSE index (Great Britain). Prediction series, calculated with different learning set's length.}
%\label{fig:Fig13a}
%\end{figure}

%\begin{figure}
%\centerline{\includegraphics[width=6.16in,height=4.35in]{13b_ftse_mean.png}}
%\caption{ FTSE index (Great Britain). Mean value and standard deviation for prediction series.}
%\label{fig:Fig13b}
%\end{figure}

\clearpage

The normalization procedure is proposed in order to compare indices and it's prediction series with different absolute values. The normalized values calculated with the following formula:
\begin{equation}
\label{eq:eq_norm}
y_n(t)=\frac{y(t)-min\left(y(t)\right)}{max\left(y(t)\right)-min\left(y(t)\right)}.
\end{equation}

\begin{figure}[hp]
\centerline{\includegraphics[width=6.16in,height=4.35in]{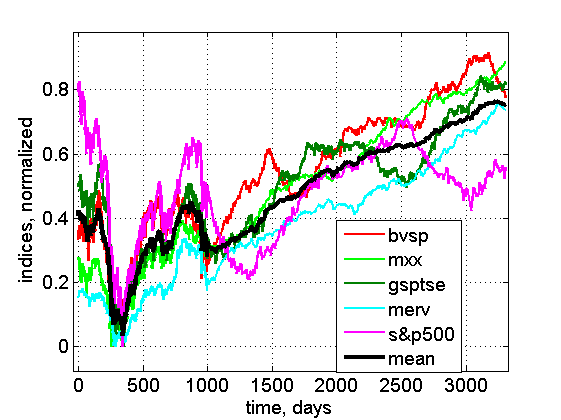}}
\caption{ Normalized mean values for the prediction series of America's stock indices. Brazil (BVSP), Mexico (MXX), Canada (GSPTSE), Argentina (MERV), USA (S\&P 500)}
\label{fig:Fig14}
\end{figure}

\begin{figure}[tbhp]
\begin{minipage}[h]{0.8\linewidth}
\includegraphics[width=1\linewidth]{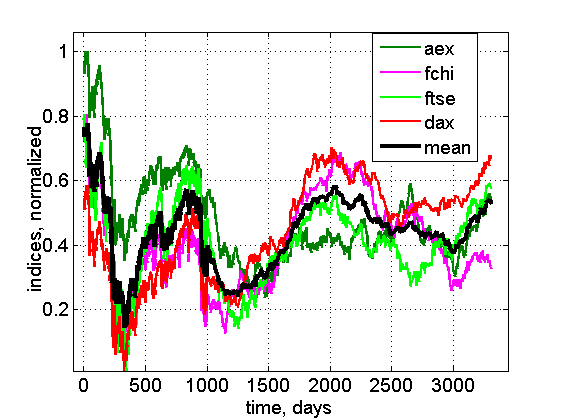} %[width=6.16in,height=4.35in]
\caption{ Normalized mean values for the prediction series of European stock indices. Developed countries: FTSE (Great Britain), DAX (Germany) FCHI (France), Netherlands (AEX)} %дания ли? Оказывается, Нидерланды. 
\label{fig:Fig15}
\end{minipage}
\vfill
\begin{minipage}[h]{0.8\linewidth}
\includegraphics[width=1\linewidth]{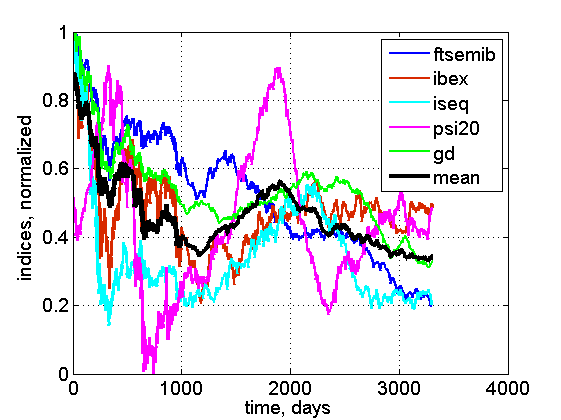} %[width=6.16in,height=4.35in]
\caption{ Normalized mean values for the prediction series of European stock indices. Portugal (PSI20), Italy (FTSEMIB), Ireland (ISEQ), Greece (GD) and Spain (IBEX).}
\label{fig:Fig16}
\end{minipage}
\end{figure}

\begin{figure}[tbhp]
\begin{minipage}[h]{0.8\linewidth}
\includegraphics[width=1\linewidth]{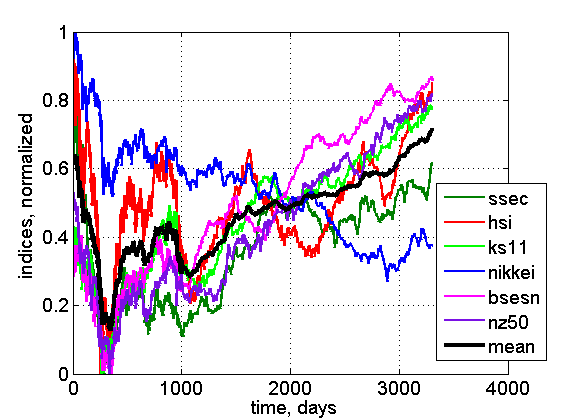} %[width=6.16in,height=4.35in]
\caption{ Normalized mean values for the prediction series of Asian stock indices. China (SSEC, HSI), Korea (KS11), Japan (NIKKEI), India (BSESN), New Zealand (NZ50). } % страны написать
\label{fig:Fig17}
\end{minipage}
\vfill
\begin{minipage}[h]{0.8\linewidth}
\includegraphics[width=1\linewidth]{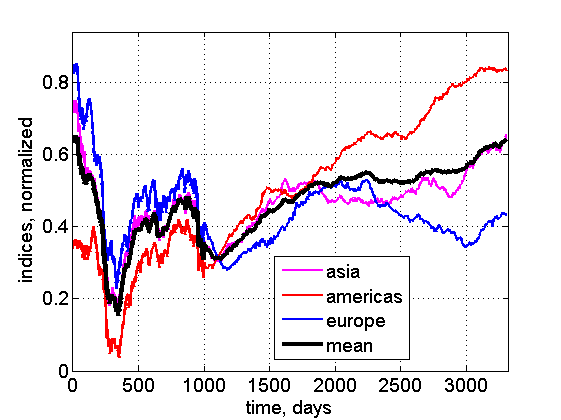} %[width=6.16in,height=4.35in]
\caption{ Mean values of normalized World's powerful economies indices prediction series.}
\label{fig:Fig18}
\end{minipage}
\end{figure}

Normalized prediction time series are shown at the fig.\ref{fig:Fig14} (America), fig.\ref{fig:Fig15} (Europe, developed countries), fig.\ref{fig:Fig16} (Europe, PIIGS), fig.\ref{fig:Fig17} (Asian markets). All the figures contain mean time series, which are weighted average of countrie's stock indices predictions, weihted with GDP values \cite{EconomyWatch_com} for the corresponding countries. 

%\begin{figure}[thp]
%\includegraphics[width=6.16in,height=4.35in]{15_Europe_developed.png}
%\caption{ Normalized mean values for the prediction series of European stock indices. Developed countries: FTSE (Great Britain), DAX (Germany) %FCHI (France), Denmark (AEX)} %дания ли?
%\label{fig:Fig15}
%\end{figure}

%\begin{figure}[bhp]
%\centerline{\includegraphics[width=6.16in,height=4.35in]{16_Europe_piigs.png}}
%\caption{ Normalized mean values for the prediction series of European stock indices. Portugal (PSI20), Italy (FTSEMIB), Ireland (ISEQ), Greece (GD) and Spain (IBEX).}
%\label{fig:Fig16}
%\end{figure}
%\clearpage

%\begin{figure}[thp]
%\centerline{\includegraphics[width=6.16in,height=4.35in]{17_Asia.png}}
%\caption{ Normalized mean values for the prediction series of Asian stock indices.} % страны написать
%\label{fig:Fig17}
%\end{figure}

%\begin{figure}[bhp]
%\centerline{\includegraphics[width=6.16in,height=4.35in]{18_World_itog.png}}
%\caption{ Mean values of normalized World's powerful economies indices prediction series.}
%\label{fig:Fig18}
%\end{figure}

\clearpage

\section{Conclusions and further work }

Current paper suggests an algorithm of time series prediction based on complex Markov chains. Hierarchy of time increments principle allows to use the information, which is contained in the time series during the prognosis construction, to its fullest. Experimental work on stock market indices time series prediction shows the efficiency of the algorithm and confirms the relevance of further research of the offered method.

%\section{References}
%\bibliographystyle{plain}
\bibliographystyle{unsrt}
\bibliography{ChabMarkovPrognozPSEP_eng}

\begin{thebibliography}{10}

\bibitem{r001SamarskyMihaylov01}
A.~A. Samarskii and A.~P. Mikhailov.
\newblock {\em Mathematical Modelling: Ideas. Methods. Examples}.
\newblock Fizmatlit, Moscow, 2001.

\bibitem{r002BogoboyashyKurbanov04}
V.~V. Bohoboyaschyy, K.~R. Kurbanov, P.B. Paly, and V.M. Shmandiy.
\newblock {\em Principles of Forecasting in Ecology: Textbook (in Ukrainian)}.
\newblock Center navchalnoji literaturi, Kyiv, 2004.


\bibitem{IvakhnenkoMGUA68}
O.~G. Ivakhnenko.
\newblock Grouping method of data handling - the concurrent of stochastic
  approximation methods (in ukrainian)
\newblock {\em Automatika}, 3(3):58--72, 1968.


\bibitem{AndersenGluzmanSornette2000}
J.~V. {Andersen}, S.~{Gluzman}, and D.~{Sornette}.
\newblock {Fundamental framework for ``technical analysis'' of market prices}.
\newblock {\em European Physical Journal B}, 14:579--601, March 2000.

\bibitem{PincakCurrencyStringTheory2011}
D.~{Horvath} and R.~{Pincak}.
\newblock {From the currency rate quotations onto strings and brane world
  scenarios}.
arXiv:1104.4716 , April 2011.

\bibitem{PincakStringPrediction2011}
M.~{Repasan} and R.~{Pincak}.
\newblock {The string prediction models as application to financial forex
  market}.
arXiv:1109.0435 , September 2011.



\bibitem{r003ElutinKrivchenko76}
P.~V. Elyutin and V.~D. Krivchenkov.
\newblock {\em Quantum Mechanics with tasks}.
\newblock Nauka, Moscow, 1976.

\bibitem{r004LandauQuantNerelyativistic}
L.D. Landau and E.M. Lifshitz.
\newblock {\em Quantum mechanics: non-relativistic theory}.
\newblock Teoreticheska fizika (Izd. 3-e) (Landau, L. D, 1908-1968).
  Butterworth-Heinemann, 1977.

\bibitem{r005SapirEconTheoryNeodnorSyst}
J.~Sapir.
\newblock {\em K Ekonomitcheskoj teorii neodnorodnyh sistem - opyt
  issledovanija decentralizovannoj ekonomiki}.
\newblock GU VShE, Moskov, 2001.

\bibitem{r006BertalanfyGenSyst62}
L.~von Bertalanffy.
\newblock General system theory---a critical review.
\newblock {\em ``General Systems''}, VII:1--20, 1962.

\bibitem{r007KurbanovSaptsin07}
K.~R. Kurbanov and V.~M. Saptsin.
\newblock Markov chains as technology for social, economic and ecological
  processes forecasting.
\newblock In {\em ``Problemy ta perspectivy rozvitku regіonalnoї rinkovoї
  ekonomіki'' Conference proceedings (in Russian)}, pages 10 -- 14, Kremenchuk,
  May, 11-13 2007.

\bibitem{r008SaptsinMarkovPaper09_}
V.~M. Saptsin.
\newblock Experience of using genetically complex {Markov} chains for the
  neural network technology forecasting.
\newblock {\em Visnyk Krivorizkogo ekonomichnogo institutu KNEU}, 2 (18):56 --
  66, 2009.

\bibitem{r009LukashinAdapt03}
Y.~P. Lukashin.
\newblock {\em Adaptive Methods of Time Series Forecasting: Textbook}.
\newblock Finance and Statistics, Moscow, 2003.

\bibitem{r010ZaychenkoMonogr08}
Y.~P. Zaichenko.
\newblock {\em Fuzzy models and techniques in intelligent systems. Monograph(in
  Russian)}.
\newblock Slovo, Kiev, 2008.

\bibitem{r011EzhovShumsky98}
A.~A. Ezhov and S.~A. Shumsky.
\newblock {\em Neurocomputing and its application in economics and business
  (Series ``Textbooks'' of Economic-Analytical Institute MEPI) / Ed. prof. V.
  V. Kharitonov}.
\newblock MEPI, Moscow, 1998.

\bibitem{r012Zaencev}
I.~V. Zayencev.
\newblock {\em Neural networks: basic models}.
\newblock {Textbook for the course ``Neural networks'' for 5-th grade students
	(in Russian) Physical Electronics Department, Faculty of Voronezh
	State University}
\newblock Voronezh State University, Voronezh.

\bibitem{r013ChabM10_}
D.~M. Chabanenko.
\newblock Detection of short- and long-term memory and time series prediction
  methods of complex {Markov} chains.
\newblock In {\em Visnyk Natsionalnogo tehnichnogo universitetu ``Kharkivsky
  politehnichny institut''. Zbirnik Naukovyh pratz. Tematichny vypusk:
  Informatika i modelyuvannya (in Ukrainian)}, number~31, pages 184 -- 190. NTU
  KHPI, Kharkov, 2010.
  \href{http://www.pim.net.ua/ARCH_F/V_pim_10.pdf}{http://www.pim.net.ua/ARCH\_F/V\_pim\_10.pdf}

\bibitem{r014SobloevFunctProstr}
S.~L. Sobolev.
\newblock {\em Selected topics from the theory of functional spaces and
  generalized functions}.
\newblock Nauka, Moscow, 1989.

\bibitem{r015HeliFunctDiffury}
J.~Hale.
\newblock {\em Theory of Functional Differential Equations}.
\newblock Springer-Verlag, New York, Heidelberg, Berlin, 1977.

\bibitem{r016BukingemShumy}
M.J. Buckingham.
\newblock {\em Noise in electronic devices and systems}.
\newblock Ellis Horwood series in electrical and electronic engineering. E.
  Horwood, 1983.

\bibitem{r017LorenzNonlinear89}
Hanz-Valter Lorenz.
\newblock {\em Nonlinear Dynamical Economics and Chaotic Motion}.
\newblock Springer-Verlag, 1989.

\bibitem{r018PetersChaosOrder}
E.E. Peters.
\newblock {\em Chaos and order in the capital markets: a new view of cycles,
  prices, and market volatility}.
\newblock Wiley finance editions. Wiley, 1996.

\bibitem{r019FederFractals}
J.~Feder.
\newblock {\em Fractals}.
\newblock Physics of solids and liquids. Plenum Press, 1988.

\bibitem{r020TihonovMironovMarkovProcesses}
V.~I. Tikhonov and V.~A. Mironov.
\newblock {\em Markov Processes}.
\newblock Soviet Radio, Moscow, 1977.

\bibitem{r021KornVMSpravochnik73}
G.A. Korn and T.M. Korn.
\newblock {\em Mathematical handbook for scientists and engineers: definitions,
  theorems, and formulas for reference and review}.
\newblock Dover books on mathematics. Dover Publications, 2000.

\bibitem{RafteryHighOrderMarkovChain}
Andrian~E. Raftery.
\newblock A model for high-order markov chains.
\newblock {\em Journal of the Royal Statistical Society.}, 1985.

\bibitem{RafteryTavare94}
Adrian Raftery and Simon Tavare.
\newblock Estimation and modelling repeated patterns in high order markov
  chains with the mixture transition distribution model.
\newblock {\em Appl. Statist.}, 43(1):179--199, 1994.

\bibitem{r022SaptsinSoloviev_}
V.~M. Saptsin and V.~N. Soloviev.
\newblock {\em Relativistic Quantum Econophysics. New paradigms of Complex
  systems modeling: Monograph}.
\href{http://kafek.at.ua/sol_sap_monogr.rar}{http://kafek.at.ua/sol\_sap\_monogr.rar}
\newblock Brama-Ukraine, Cherkassy, 2009.


\bibitem{r022SapSolArxiv_}
V.~{Saptsin} and V.~{Soloviev}. Relativistic quantum econophysics - new paradigms in complex systems  modelling. arXiv:0907.1142v1 {[}physics.soc-ph{]}.


\bibitem{r023SolDerbMonogr2010_}
V.~D. Derbentsev, A.~A. Serdyuk, V.~N. Soloviev, and O.~D. Sharapov.
\newblock {\em Synergetical and econophysical methods for the modeling of
  dynamic and structural characteristics of economic systems. Monograph (in
  Ukrainian)}.
\newblock Brama-Ukraine, Cherkassy, 2010.
\href{http://kafek.at.ua/Monogr.pdf}{http://kafek.at.ua/Monogr.pdf}

\bibitem{r024SapChFourierKharkov}
V.~M. Saptsin and D.~N. Chabanenko.
\newblock Fourier-based forecasting of low-frequency components of economical
  dynamic's time series.
\newblock In {\em Problemy ekonomichnoyi kibernetiki: Tezy dopovidey XIV
  Vseukrayinskoyi Naukovo-praktichnoyi konferentsiyi.(in Ukrainian)}, page 132,
  Kharkiv, Oct 8-9, 2009 2009. KhNU imeni VN Karazina.

\bibitem{r025ChabS10_}
D.~M. Chabanenko.
\newblock Discrete {Fourier}-based forecasting of time series.
\newblock {\em Sistemni tehnologii. Regionalny mizhvuzivsky zbirnik naukovyh
  pratz (in Ukrainian)}, 1(66):114 -- 121, 2010.
 \href{http://www.nbuv.gov.ua/portal/natural/syte/2010_1/15.pdf}{http://www.nbuv.gov.ua/portal/natural/syte/2010\_1/15.pdf}

\bibitem{r026solovievmatheconomics_}
V.~N. Soloviev.
\newblock {\em Mathematical economics. A textbook for self-study (in
  Ukrainian)}.
\newblock CHNU, Cherkassy, 2008.
\newblock{\href{http://kafek.at.ua/Posibnyk_Soloviev.rar}{http://kafek.at.ua/Posibnyk\_Soloviev.rar}}

\bibitem{r027SolSapChDrezden09}
Vladimir Soloviev, Vladimir Saptsin, and Dmitry Chabanenko.
\newblock Prediction of financial time series with the technology of high-order
  markov chains.
\newblock In {\em Working Group on Physics of Socio-economic Systems (AGSOE)},
  Drezden, 2009.
 \href{http://www.dpg-verhandlungen.de/2009/dresden/agsoe.pdf}{http://www.dpg-verhandlungen.de/2009/dresden/agsoe.pdf}

\bibitem{r028ChabRiga10}
V.~Soloviev, V.~Saptsin, and D.~Chabanenko.
\newblock Financial time series prediction with the technology of complex
  markov chains.
\newblock {\em Computer Modelling and New Technologies}, 14(3):63--67, 2010.
\href{http://www.tsi.lv/RSR/vol14_3/14_3-7.pdf}{http://www.tsi.lv/RSR/vol14\_3/14\_3-7.pdf}


\bibitem{Finance_yahoo}
Yahoo! finance database.
\href{http://finance.yahoo.com}{http://finance.yahoo.com}

\bibitem{EconomyWatch_com}
Economic indicators : Gdp per capita (current prices, national currency) 2010.
\href{http://www.economywatch.com/economic-statistics/economic-indicators/GDP_Per_Capita_Current_Prices_National_Currency/2010/}{http://www.economywatch.com/economic-statistics/economic-indicators/GDP\_Per\_Capita\_Current\_Prices\_National\_Currency/2010/}



\end{thebibliography}

\end{document}